\RequirePackage[log]{snapshot}

\documentclass[journal]{IEEEtran}

\usepackage{cite}
  \usepackage[pdftex]{graphicx}
  \graphicspath{{./Fig/}}
  \DeclareGraphicsExtensions{.pdf,.jpeg,.png}

\usepackage[cmex10]{amsmath}
\usepackage[]{amssymb}

\usepackage{algorithm}

\usepackage{algorithmic}

\usepackage{array}

 \usepackage{dblfloatfix}

\usepackage[usenames,dvipsnames]{xcolor}
\usepackage{lipsum}
\usepackage{bm}

\newcommand{\snr}{\textsf{SNR}}

\definecolor{AnnotCol}{HTML}{990000}

\newcommand{\mtight}{\setlength{\thickmuskip}{0mu} \setlength{\medmuskip}{0mu} \setlength{\thinmuskip}{0mu}}

\begin{document}

\title{Resolving Weak Sources within a Dense Array using a Network Approach}

\author{Nima Riahi$^*$ 
        and~Peter~Gerstoft \\{\em Scripps Institution of Oceanography\\University of California San Diego, La Jolla, California, 92093}
\thanks{$^*$ E-mail contact for correspondence: nriahi@ucsd.edu}
\thanks{}
\thanks{}}

%
%


\ifCLASSOPTIONpeerreview
	\markboth{}%
	{Weak Sources in Dense Arrays}
\else
	\markboth{}%
	{Riahi and Gerstoft: Weak Sources in Dense Arrays}
\fi
%



\maketitle

\begin{abstract}
A non-parametric technique to identify weak sources within dense sensor arrays is developed using a network approach. No knowledge about the propagation medium is needed except that signal strengths decay to insignificant levels within a scale that is shorter than the aperture.
We then reinterpret the spatial covariance matrix of a wave field as a matrix whose support is a connectivity matrix of a network of vertices (sensors) connected into communities.
These communities correspond to sensor clusters associated with individual sources.
We estimate the support of the covariance matrix from limited-time data using a robust hypothesis test combined with a physical distance criterion. The latter ensures sufficient network sparsity to prevent vertex communities from forming by chance.
We verify the approach on simulated data and quantify its reliability. 
The method is then applied to data from a dense 5200~element geophone array that blanketed $7\times 10$~km of the city of Long Beach (CA). The analysis exposes a helicopter traversing the array, oil production facilities, and reveals that low-frequency events tend to occur near roads.
\end{abstract}


%
\IEEEpeerreviewmaketitle

\section{Introduction}
%
%
%
%

\IEEEPARstart{L}{arge} aperture sensor arrays with dense spatial sampling are becoming more common as the cost for sensor and communications hardware decreases. Examples of such arrays are the USArray initiative in seismology with 500~stations covering large parts of the continental~US~\cite{kerr2013a} or dense seismic arrays for seismic exploration with 5200~sensors as presented here.

This work addresses the task of localizing weak sources within a fraction of a dense array. Such sources naturally become more common as the array aperture increases. 
The task is solved by considering a sensor array as a network of vertices (the sensors) where the relation between vertices is defined by high coherence and spatial proximity of the underlying sensors. For the case where sources within the array are not too close the resulting network has several disconnected components, each corresponding to a spatially contiguous sensor cluster sensing one of the sources.

Source localization with sensor arrays in complex environments has been addressed using the model-based framework of matched field processing (MFP).
The method originates in ocean acoustics~\cite{bucker1976a,baggeroer1993a,booth1996a,korakas2011a} but has found applications in seismology~\cite{harris2010a,corciulo2012a} and electromagnetics~\cite{gingras1997a,valtr2011a}. 
The structure of the covariance matrix plays an in important role in these approaches, in particular for data-adaptive implementations using, e.g., MVDR~\cite{capon1969a} or MUSIC~\cite{schmidt1986a}. But MFP methods rely on knowledge about the propagation medium in some way. If such information is not sufficiently available their applicability may be severely limited. 

Our approach assumes limited knowledge about the propagation medium. The only assumption made is that weak signals enter the noise floor within a problem-dependent distance~$d$ that is smaller than the array aperture. That requirement is realistic for large arrays in moderately attenuating media such as the earth. If the sources are not too close we will see that the covariance matrix has a structure that suggests a reinterpretation: the matrix of spatial covariance of a wave field at different sensor locations is turned into the connectivity matrix that describes the organization of a network of sensors. From that vantage point finding weak within-aperture sources becomes identifying connected sensor communities. 
The general applicability of the method comes at the cost of precision as a source is only localized through the identification of the sensors on which it has had a significant impact. By definition this set of sensors typically corresponds to a small fraction of sensors in a dense array.

We are not aware of any method that solves this source localization problem. It is conceivable that methods based on singular value decomposition of the coherence matrix or the correlation of the evolution of covariance matrix elements could also be used. These approaches are not developed here. 
Our network framework uses the covariance matrix structure to identify subspaces within that matrix that represent local sources.

In section~\ref{sec:sparse-recon-with-graph} we define the problem setting and explain how the array covariance matrix has a structure that allows it to be reinterpreted as the connectivity matrix of a network.
In section~\ref{sec:coh-spat-decay} we describe why we use coherence instead of the covariance and how the distance~$d$ can be estimated. Section~\ref{sec:find-coh-cluster} first introduces some network terminology and then describes how an appropriate network can be constructed from the coherence matrix. We verify and test the reliability of the technique on simulated data in section~\ref{sec:application-sim}. This is followed in section~\ref{sec:LBdata} by an application to real data from a 5200~sensor geophone array that was deployed in the city of Long~Beach~(CA).

\hfill August 5th, 2015

\section{Covariance Matrix Support defines a Network}
\label{sec:sparse-recon-with-graph}

Consider a large aperture array with~$N$ sensors distributed densely over spatial locations $\{ {\bf r}_i \}_{i=1,\ldots,N}$. 
We assume that there are weak sources within the aperture that produce signals that can  propagate through space. The channel between any such source location $\bm\rho$ and sensor location ${\bf r}_i$ is characterized by a Green's function, $g({\bf r}_i,{\bm \rho})$.
Let the vector ${\bf g}(\bm\rho) = [g({\bf r}_1,{\bm \rho}),\ldots,g({\bf r}_N,{\bm \rho})]^T \in \mathbb{C}^N$ be the frequency domain response of the array to a source at location ${\bm \rho}$ (equations in this article apply to the frequency domain). Consider then $\bm \rho_i$ to span a grid of~$M$ possible source locations $\{{\bm \rho}_i\}_{i=1,\ldots,M}$ with each such possible location having an associated response ${\bf g}(\bm \rho_i) \equiv {\bf g}_i$.
We define the matrix ${\bf G}=[{\bf g}_1,\ldots, {\bf g}_M]$  containing the array response to a source all of those locations.
If we let the source signals $s_i$ at those possible locations $\bm \rho_i$ be collected in the source vector ${\bf s}=[s_i,\ldots,s_M]^T \in \mathbb{C}^M$ then the measured signal at the~$N$ array sensors is modelled as:
\begin{equation}
\label{eq:model}
{\bf x} = {\bf G}\, {\bf s} + {\bf n} \: ,
\end{equation}
where ${\bf n}=[n_1,\ldots,n_N]\in \mathbb{C}^N$ is a multivariate i.i.d. noise process.
A common approach to estimate the source vector ${\bf s}$ and hence localize the sources is matched-field processing~(MFP)~\cite{baggeroer1993a}. The method has been applied to acoustic~\cite{booth1996a,korakas2011a}, seismic~\cite{harris2010a,corciulo2012a}, and electromagnetic~\cite{gingras1997a,valtr2011a} wave fields.

An important requirement for~MFP to work is the knowledge of the propagation medium encapsulated in $g({\bf r},\bm \rho)$. Consider now a grid-free formulation for the observation vector assuming a source distribution function $s(\bm\rho)$ with sources at locations $\bm\rho_1,\ldots,\bm\rho_K$:
\begin{align}
{\bf x} &= \int {\bf g}({\bm \rho})s(\bm\rho) d\bm\rho \:\: + \:\: {\bf n} \\
 &= \int {\bf g}({\bm \rho}) \left[ \sum_{i=1}^K s_i \delta(\bm\rho - \bm\rho_i)  \right] d\bm\rho \:\: + \:\: {\bf n} \\
	&= \sum_{i=1}^K s_i \: {\bf g}_i + {\bf n} \: .
\end{align}

The covariance matrix in this scenario becomes:
\begin{align}
\label{eq:modelcov1}
{\bf C} 	&= \left< {\bf x}{\bf x}^H \right> 
			= \left< \sum_{i,j=1}^K s_is^*_j\;{\bf g}_i{\bf g}^H_j \right> + \left< {\bf n}{\bf n}^H\right> \: ,   \\
			&= \sum_{i,j=1}^K \langle s_is^*_j \rangle {\bf g}_i{\bf g}^H_j + \left< {\bf n}{\bf n}^H\right> \nonumber \\ 
\label{eq:cov-is-sparse-sum}
			&= \sum_{i=1}^K |s_i|^2 \; {\bf g}_i{\bf g}^H_i  + {\bf D}_n\; ,
\end{align}

where we exploit the mutual independence between the source and noise processes $\left< s_in^*_j \right>=0, \left< n_in^*_j\right>=\left< s_is^*_j\right>=\delta_{ij}, \: \forall i,j$ and ${\bf D}_n$ is a diagonal matrix whose diagonal element $D_{ii}$ is the noise variance of sensor~$i$. 


We now assume that information about $g({\bf r},\bm \rho)$ is severely limited: we only know that the source signal at location ${\bm \rho}_i$ has no significant effect at a sensor beyond a distance~$d$, i.e. the $k$-th element ${\bf g}_i^{(k)}=0$ if $|{\bm \rho}_i - {\bf r}_k|>d$.
This is equivalent to saying that the support of~${\bf g}_i$ indicates the sensors affected by the source at location ${\bm \rho}_i$. This means that if all sources are mutually separated by at least $2d$, $|\bm\rho_i-\bm\rho_j|>2d \:\:\: \forall i\ne j$, then the corresponding support sets of the ${\bf g}_i$ do not overlap. 

Let us define the support indicator function $\mathcal{I}({\bf v})$ of a vector or matrix. 
The lack of overlap of the ${\bf g}_i$ and the support-indicator function properties (Appendix~\ref{app:indicator-ops}) allow us to write the support of the sum in (\ref{eq:cov-is-sparse-sum}) as
\begin{align}
\mathcal{I}\left(\sum_{i=1}^K |s_i|^2 \; {\bf g}_i{\bf g}^H_i \right) 
	&= \sum_{i=1}^K \mathcal{I}({\bf g}_i{\bf g}^H_i) \\
	\label{eq:cov-indicator-is-connectivity}
	&= \sum_{i=1}^K \mathcal{I}({\bf g}_i)\mathcal{I}({\bf g}_i)^T
\end{align}
This support of ${\bf C}$ has a very useful structure:
In appendix~\ref{app:connectivity-supp} we show that a network with~$N$ vertices and~$K$~connected components has a connectivity matrix that has a support that is equal or less than that of (\ref{eq:cov-indicator-is-connectivity}). Consequently, the connected components of a network with connectivity matrix $\mathcal{I}({\bf C})$ will lead us to the support of the ${\bf g}_i$ and hence to the sensor clusters that sensed the $K$~sources (the addition of the diagonal term does not affect this fact). 
In sections~\ref{sec:coh-spat-decay} and~\ref{sec:find-coh-cluster} we describe how an array network that retains this property can be constructed based on a limited number of snapshots.

\section{Coherence and its Spatial Decay}
\label{sec:coh-spat-decay}

In this section we develop a robust hypothesis test to establish the support of the covariance matrix. Using a simulation we then look into how coherence values decay as a function of receiver-pair separation. This guides us to a value for~$d$.

\subsection{Coherence-based hypothesis test}
\label{sec:coherence}

Consider a zero-mean stationary signal $u(\tau)$ sampled at intervals $\Delta t$. We define its discrete Fourier transform over a period $\mtight T_W=N\Delta t$ starting at time $t$ as:
\begin{equation}
\label{eq:DFTdef}
x_k(f,t) = \sum_{j=0}^{N-1} u_k(t+j \Delta t) w_j \,e^{{-i2\pi (j\Delta t)f}} \; ,
\end{equation}

where frequency is discretized as $f=\frac{k}{T_W}, k=0,\ldots,N/2$ and the weights $w(j)$ are used to control spectral leakage~(\cite{press2007a}). A sample estimate for the covariance in (\ref{eq:modelcov1}) can be computed using $M$~time snapshots of $x_k(f,t)$ (frequency dependence is implicit):
\begin{equation}
\label{eq:sample-cov}
C_{ij} = \frac{1}{M} \sum_{t=0}^{M-1} x_i(t)x_j^*(t) \; ,
\end{equation}

This sample estimate will be subject to the vagaries of the noise processes~$n_i$ at the receivers ${\bf r}_i$. A customary attempt to reduce the impact of this location-specific variations is to compute the coherence instead:
\begin{equation}
\label{eq:coh-samp}
\hat{C}_{ij}^s = \frac{ \frac{1}{M}  \sum_{t=0}^{M-1} {x_i(t)x_j^*(t)}}
 { 
 \sqrt{ \left( \frac{1}{M} \sum_{t=0}^{M-1} |x_i(t)|^2 \right) 
 \left( \frac{1}{M} \sum_{t=0}^{M-1} |x_j(t)|^2  \right) }  
 }
  \: .
\end{equation}
For our purposes the coherence is equally useful as the covariance because our interest lies in the support of the covariance matrix which, in the limit $M\rightarrow \infty$, is identical to that of the coherence.

A hypothesis test can be used to decide if an element $C^s_{ij}$ is sufficiently different from the the probability density function (PDF) it would have in the absence of a shared signal. Unfortunately, despite the inherent normalization the PDF of the sample coherence is not invariant to changes of the noise variance over time in $n_i$ and $n_j$, a very typical type of non-stationarity encountered in real data. We demonstrate this for the case of $\mtight M=19$ snapshots of independent noise signals $n_i(1),\ldots,n_i(M)$ and $n_j(1),\ldots,n_j(M)$ with variance being a function of snapshot~$l$, i.e. $n(l) \sim \mathcal{CN}(0,\sigma^2(l))$. Three situations are considered:

{\bf Stationary:} $\sigma_i^2(l)=\sigma_j^2(l)=1$ for all~$l$ (both signals are stationary).
{\bf Power step (overlap):} $\mtight \sigma_{i}^2(1\ldots 5)=\sigma_{j}^2(1\ldots 5)=20$ and~$\mtight \sigma_i^2(6\ldots 19)=\sigma_j^2(6\ldots 19)=1$ (both signals have high variance on initial five snapshots and then step down in power).
{\bf Power step (no overlap):} $\mtight \sigma_{i}^2(1\ldots 5)=20, \sigma_{i}^2(6\ldots 19)=1$ and~$\mtight \sigma_j^2(6\ldots 19)=20, \sigma_j^2(1\ldots 5)=1$ (first signal starts out with high variance and steps down while second signal starts with low variance and steps up).

\begin{figure}
\center
\includegraphics*[width=.9\columnwidth]{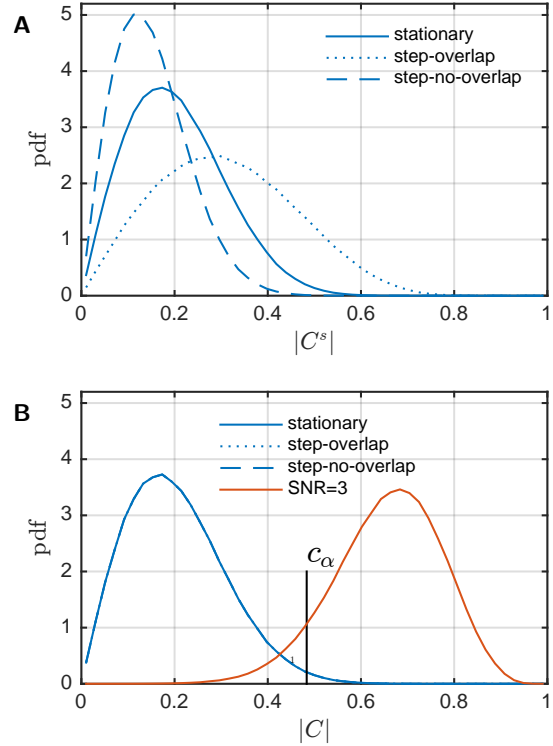} 
\caption{
	{\em (A)} The PDF of the sample coherence~(\ref{eq:coh-samp})  for independent noise-processes $n_i$ and $n_j$ whose statistics vary as a function of snapshot index according to three scenarios, two of which describe a non-stationary process (see the text for details). 
	{\em (B)} The PDF of sample coherence~(\ref{eq:coh-robust}) for the three scenarios is stable (blue lines overlap). We use this PDF to test the hypothesis that two signals share no common component. At the decision threshold $c_\alpha$=0.484 the hypothesis is falsely rejected with probability $\alpha$=0.01. The PDF of $|\hat{C}_{ij}|$ for the case of a $\snr=3$ signal being present in the noise is also shown (orange). The false acceptance rate is $\beta=0.0768$.
}
\label{fig:coh-comp}
\end{figure}

Figure~\ref{fig:coh-comp}A shows the simulated PDF of $\hat{C}_{ij}^s$ for the three scenarios (based on $10^6$ realizations). The pdf of the sample coherence~(\ref{eq:coh-samp}) substantially deviates from the stationary case for the two non-stationary scenarios considered. Such non-stationarities are, however,  common in seismo-acoustic noise processes and cannot be assumed known. A reliable hypothesis test can therefore not be based on the PDF of the sample coherence (\ref{eq:coh-samp}). We therefore consider the phase-only coherence matrix (PCM), a robust definition of the coherence~\cite{hampel2005a}:
\begin{equation}
\label{eq:coh-robust}
\hat{C}_{ij} =  
	\frac{1}{M} \sum_{t=0}^{M-1} 
		\frac{x_i(t)}{|x_i(t)|}   \frac{x_j^*(t)}{|x_j(t)|}  \: .
\end{equation}
The PCM in (\ref{eq:coh-robust}) ignores magnitudes and only relies on phase information. Figure~\ref{fig:coh-comp}B shows the PDF of $\hat{C}_{ij}$ for the same scenarios as before (blue lines) and demonstrates how the distribution of this statistic is invariant for the types of non-stationarities. We use the coherence~(\ref{eq:coh-robust}) to test this hypothesis: "the signals observed at locations ${\bf r}_i$ and ${\bf r}_j$ based on $M=19$ snapshots are uncorrelated". The hypothesis is accepted if $|\hat{C}_{ij}|\le c_\alpha$ and rejected otherwise. The threshold coherence magnitude $c_\alpha$ is set such that the probability of falsely rejecting the hypothesis is $\alpha$, formally:
\begin{equation}
\label{eq:coh-ths}
c_\alpha = \text{cdf}^{-1}(1-\alpha)  \: ,
\end{equation}
where $\text{cdf}^{-1}(\cdot)$ is the inverse of the cumulative distribution function estimated from the simulation in Figure~\ref{fig:coh-comp}B. To provide an idea about the likelihood of falsely {\em accepting} the null-hypothesis Figure~\ref{fig:coh-comp}B also shows the simulated pdf of $|\hat{C}_{ij}|$ for the case where there is a signal present with $\mtight \snr=3$. Table~\ref{tab:dec-ths} gives the decision threshold $c_\alpha$ and  false acceptance rate~$\beta$ given the false rejection rate~$\alpha$. For the remainder of this article we use coherence as a short-hand for the magnitude of the phase-only coherence  \eqref{eq:coh-robust}.

\begin{table}[]
\centering
\caption{Coherence magnitude decision threshold $c_\alpha$ based on simulation for  confidence levels $\alpha$ (i.e. false rejection probability). The false acceptance probability~$\beta$ is given for the case $\snr$=3.}
\label{tab:dec-ths}
\begin{tabular}{|c|c|c|} \hline
$\alpha$ & $c_\alpha$ & $\beta$ ($\snr=3$)  \\ \hline
0.010 & 0.484 & 0.0768   \\
0.005 & 0.517 & 0.118   \\
0.001 & 0.582 & 0.247 \\ \hline
\end{tabular}
\end{table}

\subsection{Spatial decay of coherence}
\label{sec:coh-vs-distance}

\begin{figure}
\center
\includegraphics*[width=.8\columnwidth]{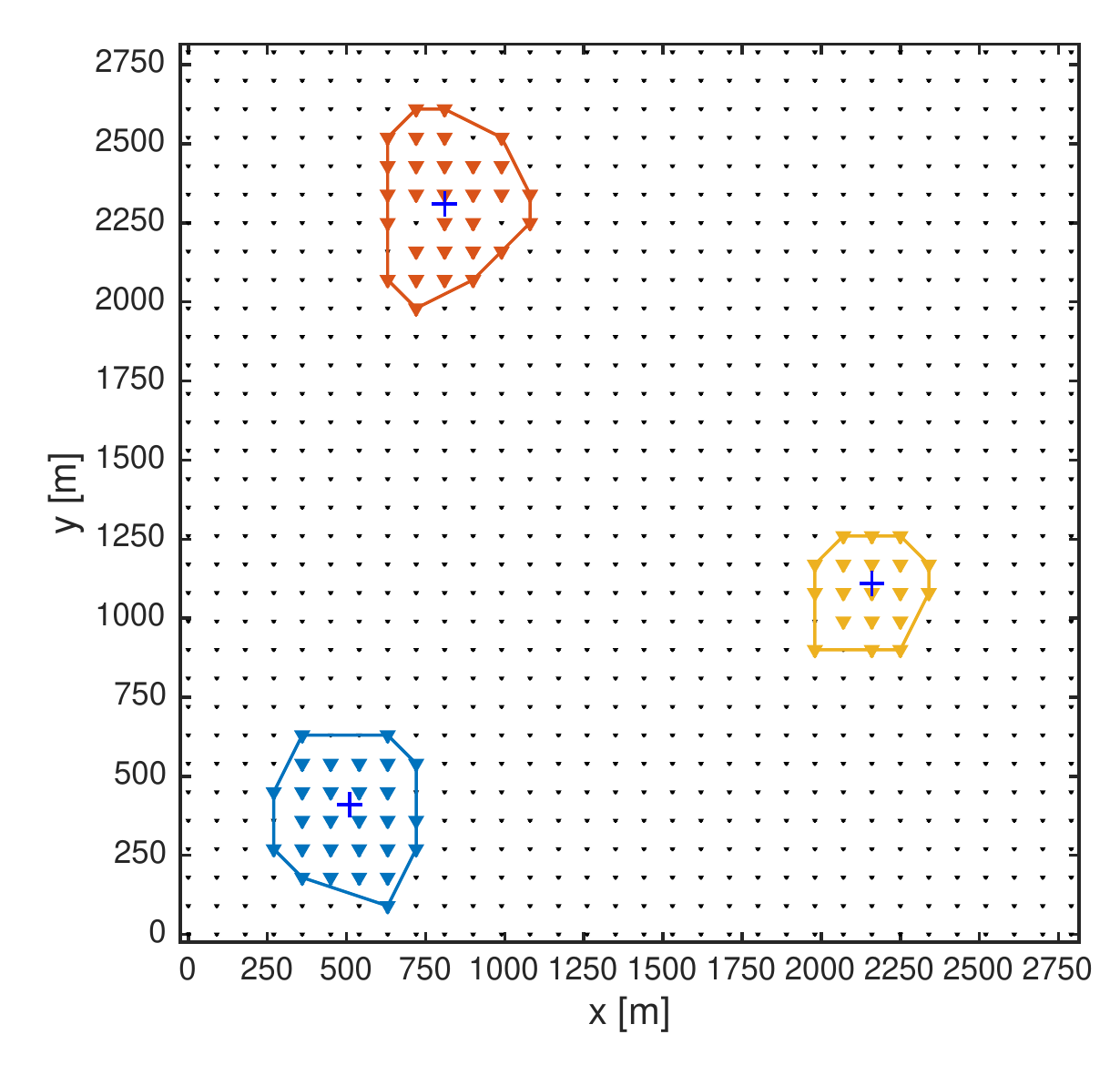}
\caption{
	Simulation layout of a 2.8$\times$2.8~km array with 1024~sensor spaced 90~m apart (dots) with three sources (blue cross, see text for details). A set of 19~snapshots was simulated with $\snr=200$ at 10~m distance from each source. The colored triangles are sensor groups representing connected subgraphs from the simulation. The colored enclosures show the convex hull of each sensor clusters.
	}
\label{fig:synth-map}
\end{figure}

The null-hypothesis rejection rate should be higher than $\alpha$ for a sensor pair in the proximity of a source. If fact, for sources with a local imprint this excess rejection rate depends on receiver pair separation. The following simulation demonstrates this.  Consider a square array on a homogeneous half-space with aperture 2.8$\times$2.8~km and sensors placed on a rectangular grid spaced 90~m, i.e. $K=1024$~sensors at positions ${\bf r}_1,\ldots,{\bf r}_K$ (Figure~\ref{fig:synth-map}). We simulate three source signals $s_1(\tau),s_2(\tau),s_3(\tau)$ at locations $\bm\rho_1,\bm\rho_2,\bm\rho_3$ (blue crosses) which spread spherically across the array at velocity $\mtight v=340$~ms$^{-1}$, generating  time-domain sensor data at location ${\bf r}_k$:
\begin{equation}
\label{eq:propmodel}
x_k(\tau) = \frac{1}{|{\bf r}_k-{\bf r}_s|} \sum_{i=1}^{3} s_i\left(\tau-\frac{|{\bf r}_k-{\bm \rho}_{i}|}{v} + \epsilon \right) + n(\tau)\; ,
\end{equation}
where $\mtight s_i(\tau){\sim} \mathcal{N}(0,\sigma^2)$, $\mtight n(\tau){\sim} \mathcal{N}(0,\sigma_n^2)$, $\mtight s_1 \perp s_2 \perp n$ are mutually uncorrelated.  An arrival time perturbation $\epsilon \sim \mathcal{N}(0,\sigma_\epsilon^2)$ with $\sigma_\epsilon$=30~ms is introduced to emulate the situation where precise knowledge of the propagation channel is lacking, e.g. due to meteorologic convections in acoustics or unknown heterogeneities in seismic data. The simulation assumes $\mtight \sigma^2/\sigma_n^2=\snr=200$ at 10~m from the source and sampling period of $\mtight \Delta t=4$~ms. Fourier coefficients $X_k(f,t)$ are computed according to (\ref{eq:DFTdef}) with $\mtight T_W=256\Delta t\simeq 1.02$~s at 20.51~Hz.

\begin{figure}
\center
\includegraphics*[width=.9\columnwidth]{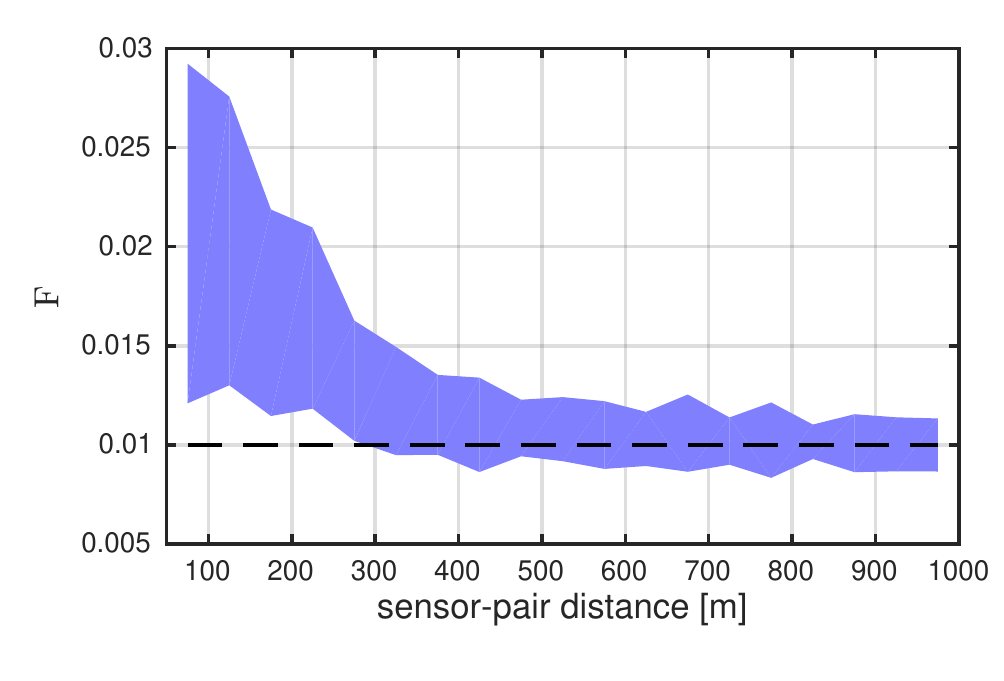} 
\caption{
	The sensor pairs $i,j$ of the simulated array are grouped into 50~m distance bins from 50--1000~m based on $d_{ij}=|{\bf r}_i-{\bf r}_j|$.
	 For each distance bin we calculate the fraction $F$ of sensor pairs where the random wave field hypothesis is rejected ($|C_{ij}|>c_\alpha$, with $\alpha=0.01$). 
	The shaded area indicates the 10--90 percentile range of these fractions from 1710~independent simulations. 
}
\label{fig:coh-vs-distance-sim}
\end{figure}

A sample coherence matrix $\hat{C}_{ij}(f)$ \eqref{eq:coh-robust} is calculated from $\mtight M=19$  snapshots of $x_k$ (\ref{eq:coh-robust}). The receiver-pairs are grouped into 50-m distance bins from 50--1000~m with each bin containing between 3.5$\cdot 10^3$ to 30$\cdot 10^3$ sensor pairs. In each group we compute the fraction of pairs that reject the null hypothesis ($|C_{ij}|>c_\alpha$) divided by the total number of pairs:
\begin{equation}
\label{eq:Hreject-fraction}
F = \frac{\text{\#}\{i,j | \: |\hat{C}_{ij}|>c_\alpha\}}{\text{\#}\{i,j\}}  \; ,
\end{equation}
where~\# refers to the cardinality of the set.
We repeat the process for~1710 simulation runs, giving us 1710~fractions for each distance bin. Figure~\ref{fig:coh-vs-distance-sim} gives the 10--90 percentile range of estimated fractions. The fractions decrease systematically with distance and for large separation are centered on the probability of falsely rejecting the random wave field hypothesis, $\alpha=0.01$ (dashed line). For receiver pairs with large separation the wave field (\ref{eq:propmodel}) has decayed so much that the fraction is not different from the expected value for a random wave field.
For distances less than 300~m, however, the observed fraction is above $\alpha=0.01$. 

\section{Constructing a Network from Array Data}
\label{sec:find-coh-cluster}

In section \ref{sec:sparse-recon-with-graph} we indicated that the structure of the support of the covariance matrix can be used to find coherent clusters using a network analysis approach. In this section we describe how an appropriate network can be constructed to find the targeted sensor clusters. We also provide an estimate on the false detection rate of the approach. We start out by introducing some terminology from network theory.


\subsection{Networks, connectedness, random networks}
\label{sec:define-graph}

An undirected and unweighted network~$G$ consists of a set of vertices $\{v_i\}_{i=1\ldots N}$ and edges $\{e_{ij}\}_{j\ge i} \in \{ 0,1 \}$ where $e_{ij}=1$ means that $v_i$ and $v_j$ are connected. The edges define a symmetric connectivity matrix~$\mtight E_{ij}=E_{ji}=e_{ij}$. The number of edges leading to vertex~$v_i$ is its degree $d_i$. 
The mean vertex degree $\gamma$ of a network is the average over the vertex degrees of all its vertices. If $\gamma$ tends to a constant as~$N$ increases the network is sparse~\cite{newman2003a}.
The notation~$|G|$ refers to the number of vertices~$N$ in network $G$. 

A connected component $U \subset G$ is a subset of vertices and edges in~$G$ for which every pair of vertices $n,n'\in S$ is connected directly or indirectly through a sequence of edges in~$S$. Finding connected components is a basic task in network analysis~\cite{newman2010a}.
Let $S$ be the only connected component in $G$ and let ${\bf u}=[u_1,\ldots,u_N]^T$ be the vertex indicator vector of~$U$ where $u_i=1$ if $i\in U$ and 0~otherwise. It can be easily verified that 
\begin{equation}
\label{eq:indicator-connectivity}
({\bf E})_{ij}=({\bf uu}^T)_{ij} = u_iu_j
\end{equation}
is the connectivity matrix of that network when~$S$ is fully connected, i.e. all vertex pairs in~$U$ are connected. From all the possible edge configurations that will leave $U$ connected this is the one with the most edges. Therefore the support of the connectivity matrix of any connected component containing the vertices in~$U$ will have the same or a smaller support than~${\bf E}$.
If we were to connect a second subset $V \subset G$ with indicator vector ${\bf v}$ that does not share vertices with $U$, $\mtight V \cap U=\emptyset$, the resulting connectivity matrix would be simply the sum ${\bf uu}^T + {\bf vv}^T$ (appendix~\ref{app:connectivity-supp}).
It follows by induction that if there are several connected subsets $U_i$ in $G$ for which $U_i \cap U_j = \emptyset \: , \forall i \ne j$ the resulting connectivity matrix is
\begin{equation}
\label{eq:conn-matrix-adding}
{\bf E}= \sum_i {\bf u}_i{\bf u}_i^T  \: ,
\end{equation}
with ${\bf u}_i$ the indicator vector of subset $U_i$. As before, any edge configuration where all subsets remain connected will have the same or a smaller support than that of ${\bf E}$.

\begin{figure}
\center
\includegraphics*[width=.8\columnwidth]{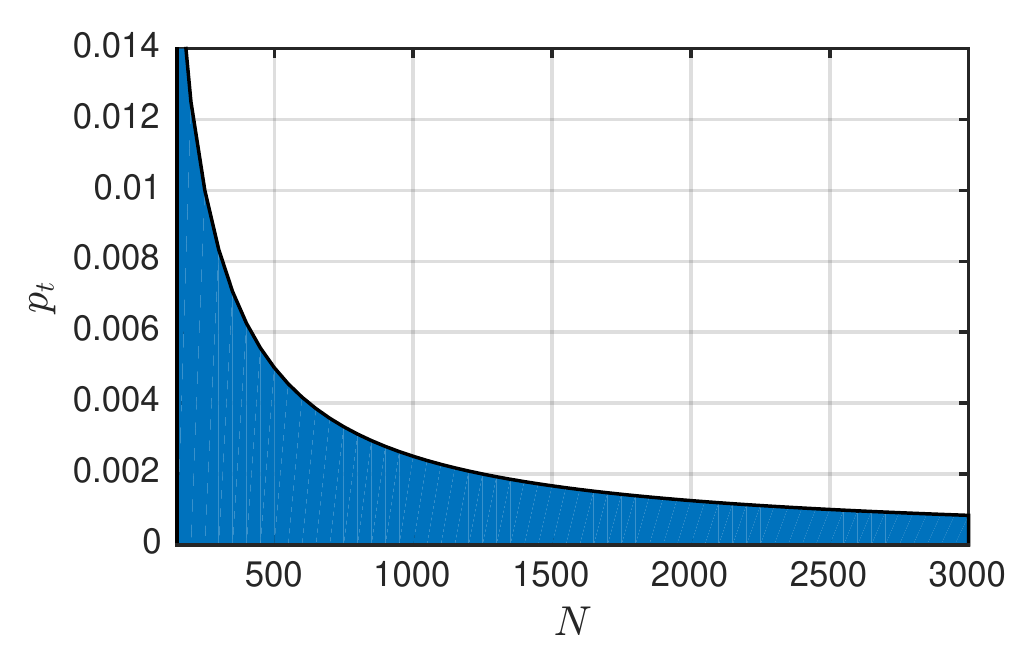}
\caption{
	The transition probability $p_t$ where most of the vertices in a random network $G_0(N,p>p_t)$, will be connected.
	}
\label{fig:conncomp-vs-Ngraph}
\end{figure}

Finally, consider the random, unweighted network $G_0(N,p)$ with~$N$ vertices where all pairs of vertices have the same probability~$p$ of being connected. The mean vertex degree in $G_0(N,p)$ is therefore $\gamma=(N-1)p$ because every vertex can connect with all $N-1$ other vertices with equal probability. A large fraction of vertices in a random network tend to be connected when the mean degree~$\gamma$ exceeds one~\cite{newman2003a}, with about 90\% being connected when $\gamma>2.5$. This means that for a connection above the transition probability $p_t=2.5/(N-1)$ most vertices will be connected. Figure~\ref{fig:conncomp-vs-Ngraph} shows the hyperbolic relation in which the transition probability $p_t$ decreases as the network grows larger. Note that random networks are not sparse since their mean vertex degree depends on the network size~$N$.


\subsection{Constructing an array network}
\label{sec:constr-graph}

We saw in (\ref{eq:cov-indicator-is-connectivity}) that the support indicator of the covariance matrix is a sum of~$K$ outer products of non-overlapping support vectors with themselves (with all diagonal elements equal to one). From (\ref{eq:conn-matrix-adding}) we now see that this corresponds exactly to the connectivity matrix of a network with~$N$ vertices and~$K$ fully connected components. 

The indicator vector of the $i$-th component would be identical to the support indicator function $\mathcal{I}({\bf g}_i)$. Finding the sensor clusters affected by the $K$~sources is thus equivalent to finding the connected components of a network with connectivity matrix $\mathcal{I}({\bf C})$ (note that the covariance and coherence matrices share the same support).
Armed with this insight we will now construct a coherence network $G^0$ with the following connectivity matrix:
\begin{equation}
\label{eq:array2graph0}
    E^0_{ij} = \begin{cases}
               1               & \text{if } |\hat{C}_{ij}| > c_\alpha  \\
               0 & \text{otherwise}
           \end{cases}
\end{equation}

The support of the covariance is thus estimated based on the hypothesis test formulated in section~\ref{sec:coherence}.
This straight-forward construction of an array network, however, is insufficient because of the statistical fluctuations of the hypothesis test.
In a random wave field the probability of observing $|\hat{C}_{ij}|>c_\alpha$ is~$\alpha$ for all receiver pairs and so in this case~$G^0$ is in fact a random network $G_0(N,\alpha)$ as defined in the previous section. As discussed there $G^0$~will tend to be connected for reasonable values of $\alpha$, in particular if the array has more than a few hundred vertices (see Figure~\ref{fig:conncomp-vs-Ngraph}). If all vertices are connected then any attempt to find smaller connected components due to a physical source is rendered futile.
The possible remedy of decreasing~$\alpha$ is not appropriate: the value cannot be made arbitrarily small because the corresponding hypothesis test would become overly conservative with a high false acceptance rate~$\beta$ (see Table~\ref{tab:dec-ths}).

We now slightly modify (\ref{eq:array2graph0}) to define the localized coherence network $G(c_\alpha,d_\text{max})$:
\begin{equation}
\label{eq:array2graph}
    E_{ij} = \begin{cases}
               1               & \text{if } |C_{ij}| > c_\alpha \: \text{and} \: |{\bf r}_i-{\bf r}_j|\le d_\text{max} \\
               0 & \text{otherwise.}
           \end{cases}
\end{equation}
We thus require not only that the coherence between any two sensors must be sufficient to reject the random signal hypothesis but also that the distance between the sensor locations must be less than~$d_\text{max}$. By design the vertices of a connected component of~$G$ represent a collection of sensors that are geographically close and share a common signal component.

Enforcing spatial localization of connections in the network limits the number of potential neighbors a vertex can connect to from the set of all vertices to just the set of vertices associated with spatially nearby sensors. This number is essentially invariant to the size of the array and can be controlled via~$d_\text{max}$. The network is therefore sparse for large arrays.

Note that this does not preclude sensor clusters with a spatial extend of more than $d_\text{max}$: as long as the true area of increased coherence is contiguous in space (at the scale of $d_\text{max}$) its full extend should be recoverable through indirect links in the connected component. To characterize the spatial extent of each such connected component~$S_j$, a mean and covariance matrix for a two-dimensional Gaussian probability density function can be estimated from the sensor locations represented by the vertices of~$S_j$:
\begin{align}
\label{eq:spat-gauss}
{\bf m}_j &= \frac{1}{|S_j|} \sum_{{\bf r}_i \in S_j} {\bf r}_i  \\ 
{\bf \Sigma}_j &= \frac{1}{|S_j|} \sum_{{\bf r}_i \in S_j} ({\bf r}_i-{\bf m}_j)({\bf r}_i-{\bf m}_j)^T 
\end{align}

These values should not be confounded with actual location estimates. Source directionality, physical obstacles or attenuation heterogeneities in the propagation medium can cause clusters that are not centered around a source. Sensor geometry, gaps in the array, and boundaries will also cause a cluster to move away from its source. The sensor cluster does, however, provide an approximate indication about source location. It can also serve as a data subset selection for follow-up analysis with more precise array processing since by definition its sensors contain significant signal levels from a given source.

\subsection{Random size of connected components}
\label{sec:conn-comp-null-distr}

\begin{figure}
\center
\includegraphics*[width=.8\columnwidth]{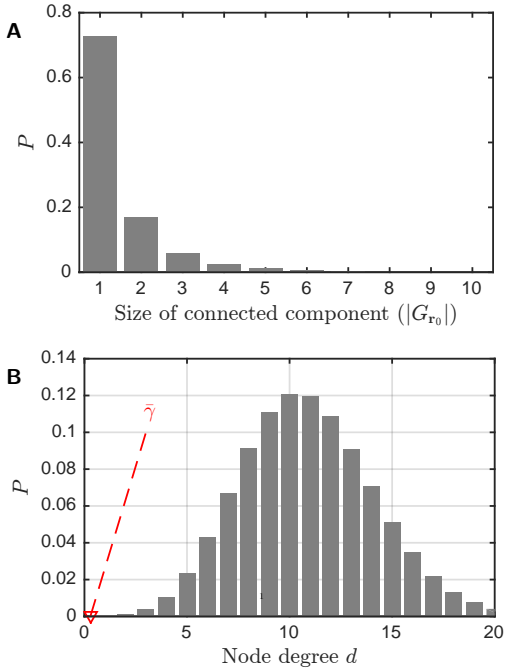}
\caption{
	Estimated probability of the size of connected components $G_{r_0}$ through reference vertex $r_0$ from $10000$ random trials {\em (A)}. The array network is constructed using $c_{0.01}=0.49$ and $d_\text{max}=300\text{~m}$. Only six trials led to connected components with $|G_{{\bf r}_0}|\ge 10$. The analytical probability distribution of vertex degrees for a random network $G_0(N,\alpha=0.01)$ is shown in {\em (B)}. The mean vertex degree~$\gamma$ of the localized coherence network $G(c_\alpha,d_\text{max})$ is shown for comparison.
	}
\label{fig:conn-comp-size-pd}
\end{figure}

To demonstrate how the distance constraint in (\ref{eq:array2graph}) prevents connected components to form by chance even in large arrays we look at the 1089~sensor layout of the simulation discussed in section~\ref{sec:coh-vs-distance} without signal sources. The probability distribution of the size of connected components is estimated for 10000~realizations of a random wave field. First, the sample coherence matrix $|\hat{C}_{ij}|$ from $M=19$ random snapshots from all sensor vertices is computed and then the corresponding array network $\mtight G(c_{\alpha=0.01},d_\text{max}=300\text{ m})$ is constructed and the size of the connected component containing a reference vertex in the center of the array, $|G_{{\bf r}_0}|$, is then stored. The simulated probability distribution over $|G_{{\bf r}_0}|$ is shown in Figure~\ref{fig:conn-comp-size-pd}A, which illustrates that most connected components in a random array will have a very small size. Of the 10000~trials there were six cases where $|G_{{\bf r}_0}|\ge 10$.
Finding a coherent sensor cluster with~10 or more vertices is therefore highly improbable to occur by chance.

The mean vertex degrees of the 10000~localized coherence networks $G(c_\alpha,d_\text{max})$ were averaged and we found $\mtight \bar{\gamma}=0.2883$. This value is compared in Figure~\ref{fig:conn-comp-size-pd}B with the theoretical probability distribution of the vertex degree for a random network $G(N,\alpha)$ of same size. Clearly, when no spatial localization is enforced the mean degree for the given array will be far above the threshold~$\mtight \gamma=1$ where the network will connect.

\section{Verification on Simulated Data}
\label{sec:application-sim}

\begin{figure}
\center
\includegraphics*[width=.9\columnwidth]{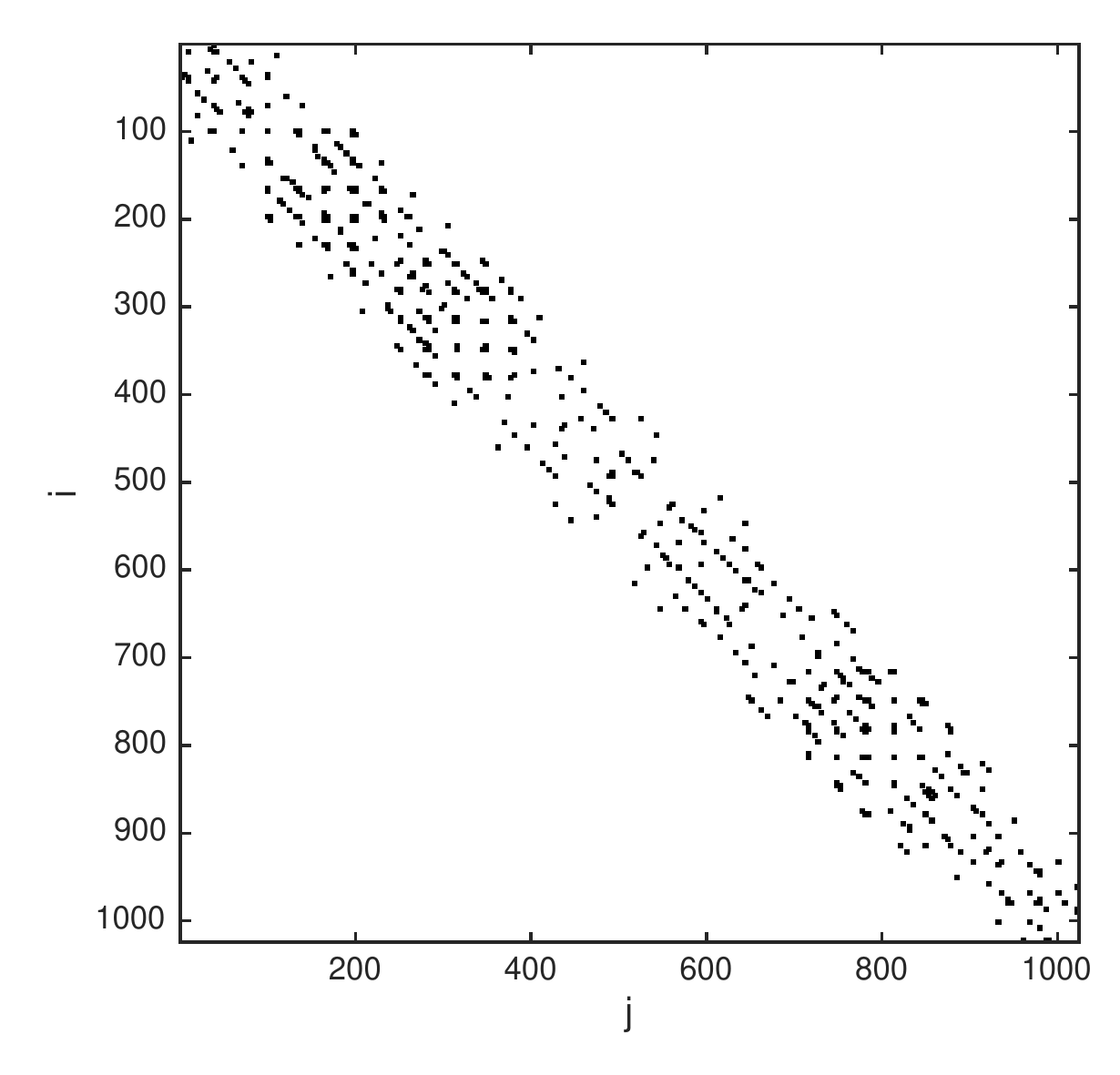} 
\caption{
The connectivity matrix for 1024~sensor array simulation. This matrix is used to make the detections in Figure~\ref{fig:synth-map}. The markers on the non-zero elements are enlarged for better visibility.
}
\label{fig:graph-conn-sim}
\end{figure}

We again return to the simulation described in section~\ref{sec:coh-vs-distance}. As discussed there, a sample coherence matrix is computed for every $M$~snapshots and coherence magnitudes $|\hat{C}_{ij}|$ are found to hover around $c_\alpha$ for distances higher than about 300~m. For each set of snapshots we therefore construct a network $\mtight G(c_{\alpha=0.01}, d_\text{max}=300\text{ m})$ and find its connected components with more than ten vertices.
Figure~\ref{fig:synth-map} shows the array configuration with the locations of the sources and the sensor clusters found in one sequence of snapshots. Figure~\ref{fig:graph-conn-sim} shows the connectivity matrix for the constructed network $\mtight G(c_{\alpha=0.01}, d_\text{max}=300\text{ m})$. It illustrates the sparsity of the network.

The performance of the approach is quantified as follows: for each simulation we count how many of the sources were not enclosed in the convex hull of a detected cluster (missed sources, i.e. false negatives) and at the same time how many of the detected clusters do not contain a source. Within 300~simulation runs (i.e. 900 sources to be detected) we found 27~missed sources (3.0\%) and 37~spurious clusters (4.1\%). A visual analysis of the simulation results (not shown) reveals that the spurious clusters where always associated with a nearby twin cluster that encompassed a source. These clusters therefore were probably caused by the presence of the source and do not correspond to the false positive rate established in section~\ref{sec:conn-comp-null-distr}. The observations indicate that a single source may occasionally give rise to more than one component around the source location. There were 922~clusters that were identified in the 300~simulation runs with a mean size of 24~vertices, indicating how small the relative size of the clusters is with respect to the array (2.3\%). 

We define a boundary for the clusters using the 1-$\sigma$ probability area of the PDF of group~$j$~(eq.~\ref{eq:spat-gauss}):
\begin{equation}
\label{eq:gauss-equiprob}
\Omega_j = \{ {\bf r} \: | \: ({\bf r}-{\bf m}_i)^T{\bf \Sigma}_i^{-1}({\bf r}-{\bf m}_i) < \chi^2_I(0.69) \} \; ,
\end{equation}
where $\chi_I^2$ is the cumulative inverse $\chi^2$\hbox{-}distribution, i.e. $E_j$ contains 69\% of the probability mass of the spatial PDF computed from the sensor locations of group~$j$.

\section{Long Beach (CA) Geophone Array}
\label{sec:LBdata}

\begin{figure*}[t]
\center
\includegraphics[origin=c,width=1.6\columnwidth]{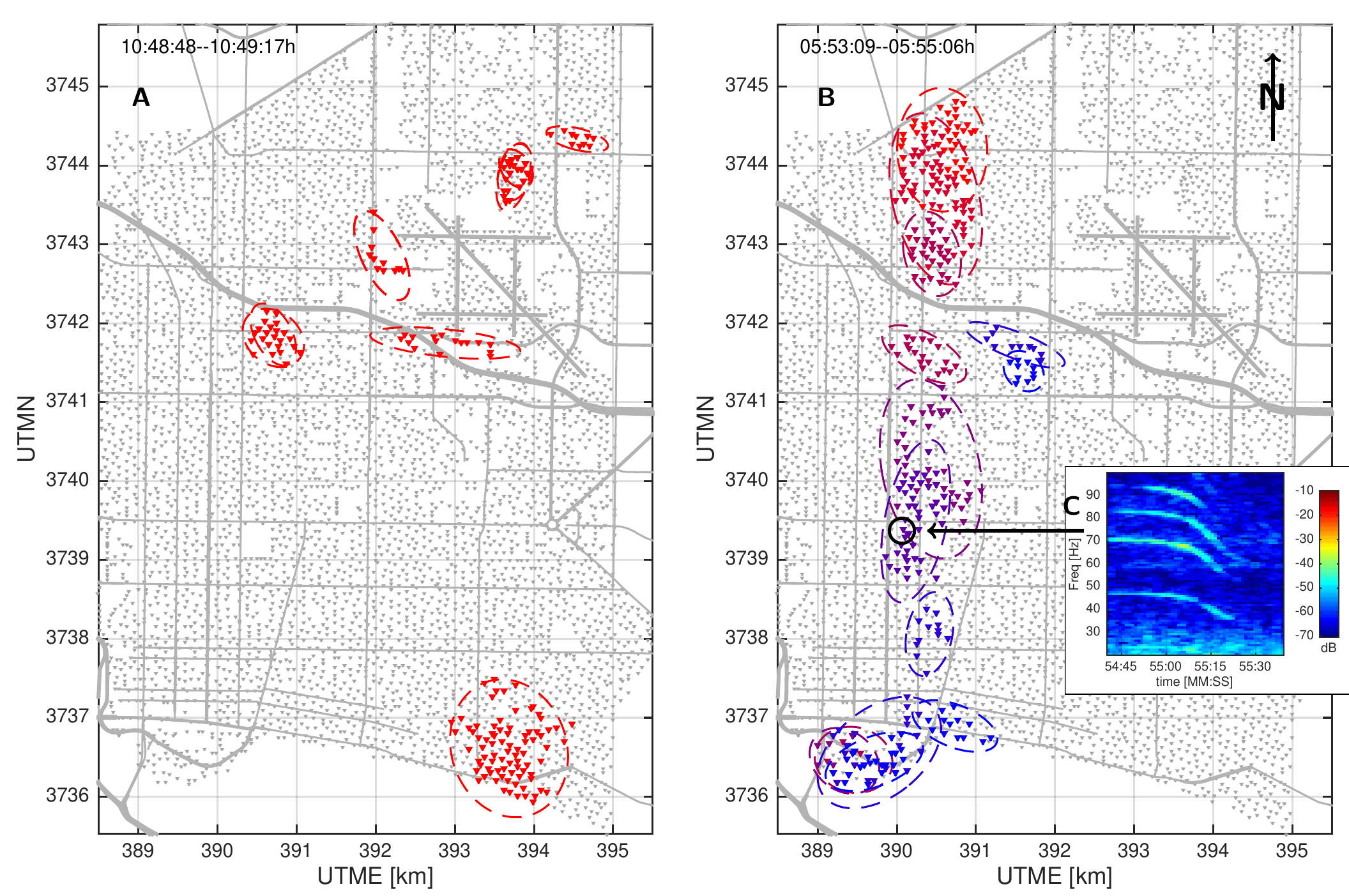}
\caption{
	Connected components of the array network were used to find coherent sensor clusters in the Long Beach geophone array. The clusters from four 9.7~s windows after 10.48h on March~11th are shown in {\em (A)}. The Southern sensor group coincides with vibrotruck activity in the Southeastern part of the array as part of the seismic acquisition operations. The dashed ellipses show the spatial extent~$\Omega$ of the groups~(\ref{eq:gauss-equiprob}).
	{\em (B)} A helicopter fly-by is captured in a sequence of coherent groups at 47~Hz over consecutive analysis time periods (starting at 05.53h, each time segment is 9.7~s advanced). The colors change from red to blue as the analysis windows advance in time. The arrow points to the receiver from which the spectrogram {\em (C)} was computed about 15~s after the coherence at that location was observed. Narrow-band harmonics at multiples of 12~Hz imply a helicopter as the source and the observed Doppler shift of $f_\text{high}/f_\text{low}\simeq 1.4$ is roughly consistent with a velocity of about 60~m/s.
    }
\label{fig:map-groups}
\end{figure*}

We demonstrate the technique  on a geophone array that was deployed over an area of about $7\times 10$~km in Long Beach (California, US) as part of an industrial seismic survey\cite{lin2013a,schmandt2013a,riahi2015a}. The array consisted of about~5200~geophones (OYO CT32D vertical velocity sensors with 10~Hz corner frequency) sampled at $\Delta t$=4~ms (see Figure~\ref{fig:map-groups}A). The sensor configuration is such that every sensor has on average one neighboring sensor within a perimeter of 90~m. We transform the velocity measurement data stream of each geophone into a sequence of Fourier coefficients $x_k(f,t)$ using $N=256$ samples ($\mtight T_W=1.02$~s) and a Hanning window $w_j$ (\ref{eq:DFTdef}) with  50\% overlap.
Linear trends in the segments are removed before computing $x_k$.
For our analysis we compute the coherence matrix $\hat{{\bf C}}_{ij}$ on 41~frequency bins from 9.8\hbox{-}48.8~Hz using $\mtight M=19$ snapshots (9.22~s). A matrix with about $\mtight 5200^2 \approx 27\cdot 10^6$~entries is therfore computed for every frequency bin and time period. In a 24~hour analysis period there are about~9400~time windows.

\begin{figure}
\center
\includegraphics*[width=.9\columnwidth]{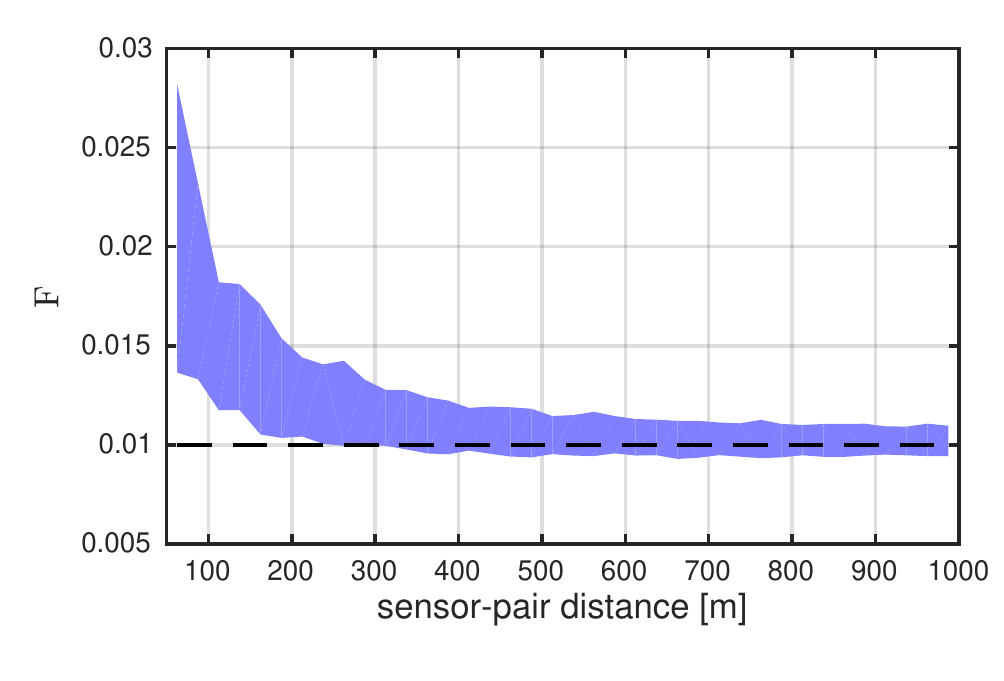} 
\caption{
Fraction $F$ of sensor pairs for which the null hypothesis is rejected  at $\alpha$=0.01 (\ref{eq:Hreject-fraction}) versus sensor separation $d_{ij}=|{\bf r}_i-{\bf r}_j|$. The shaded area indicates  the 10--90 percentiles of  fractions during 12~hours from March~10th 08.00--20.00h. Below about 300~m sensor separation there is an increasing fraction of  pairs whose coherences cannot be explained by false rejections of the random wave field hypothesis.
}
\label{fig:coh-vs-distance}
\end{figure}

As with the simulation in section~\ref{sec:coh-vs-distance}, we first study the distance dependence of the fraction~(\ref{eq:Hreject-fraction}), i.e. how much the null hypothesis rejection rate is above the background level $\alpha$. For time segments consisting of~$M=19$ snapshots the sample coherence for all receiver pairs is computed at 20~Hz following~(\ref{eq:coh-robust}). The receiver pairs are then grouped by geographic distance $d_{ij}=|{\bf r}_i-{\bf r}_j|$ into bins of width 25~m. The process is repeated for consecutive time segments from March~10th 08.00--20.00h (all time indications are local time). Figure~\ref{fig:coh-vs-distance} shows the fraction of receiver pairs in each distance group for which $|C|_{ij}>c_\alpha=0.484$ (Table \ref{tab:dec-ths}), where we set $\alpha=0.01$. We find that given the coherence threshold $c_\alpha$ it makes sense to search for coherent receiver groups within a search radius of $d_\text{max}$=300~m to avoid all the false positives associated with larger distance sensor pairs. This observation was repeated for several other frequency bins in the 10\hbox{-}50~Hz band. We find in this dataset that the value of $d_\text{max}$ depends only weakly on frequency but in general it will be problem dependent.

Having established $d_\text{max}=300$~m we  search for coherent receiver clusters in the data. We compute a coherence matrix $\hat{C}_{ij}$ at 20.0~Hz from $M=19$ consecutive snapshots which corresponds to an analysis window of 9.7~s. A localized array network~$G(c_\alpha,d_\text{max})$ is defined and all connected components with more than 10~vertices are identified. Figure~\ref{fig:map-groups}A shows the coherent groups found over four consecutive 9.7~s analysis windows starting on March~11th, 10:48:48h. The period contains a 40~s stretch during which a seismic vibrotruck was operating in the Southeast of the array, which is confirmed by a cluster of coherent receivers in that area.

Figure~\ref{fig:map-groups}B shows a sequence of coherent groups at 47~Hz for consecutive windows starting March~11th at 05.53h. They show a north-south motion over 6~km during the course of about 95~s. The corresponding velocity 60~m/s (134 mph) on the observed trajectory implies an aircraft moving across the array. Figure~\ref{fig:map-groups}C shows a spectrogram from a receiver within the trajectory of the moving source computed about 15~s after the coherence was observed. The observed Doppler shifts of $f_\text{high}/f_\text{low}\simeq 1.4 \simeq (1+\frac{60}{340})/(1-\frac{60}{340})$ are consistent with the approximate velocity estimate. Finally, the narrow-band harmonics at multiples of 12~Hz suggest that the passage of a helicopter was captured.



\subsection{Statistics over frequency and time}

\begin{figure}[]
\center
\includegraphics[origin=c,width=.9\columnwidth]{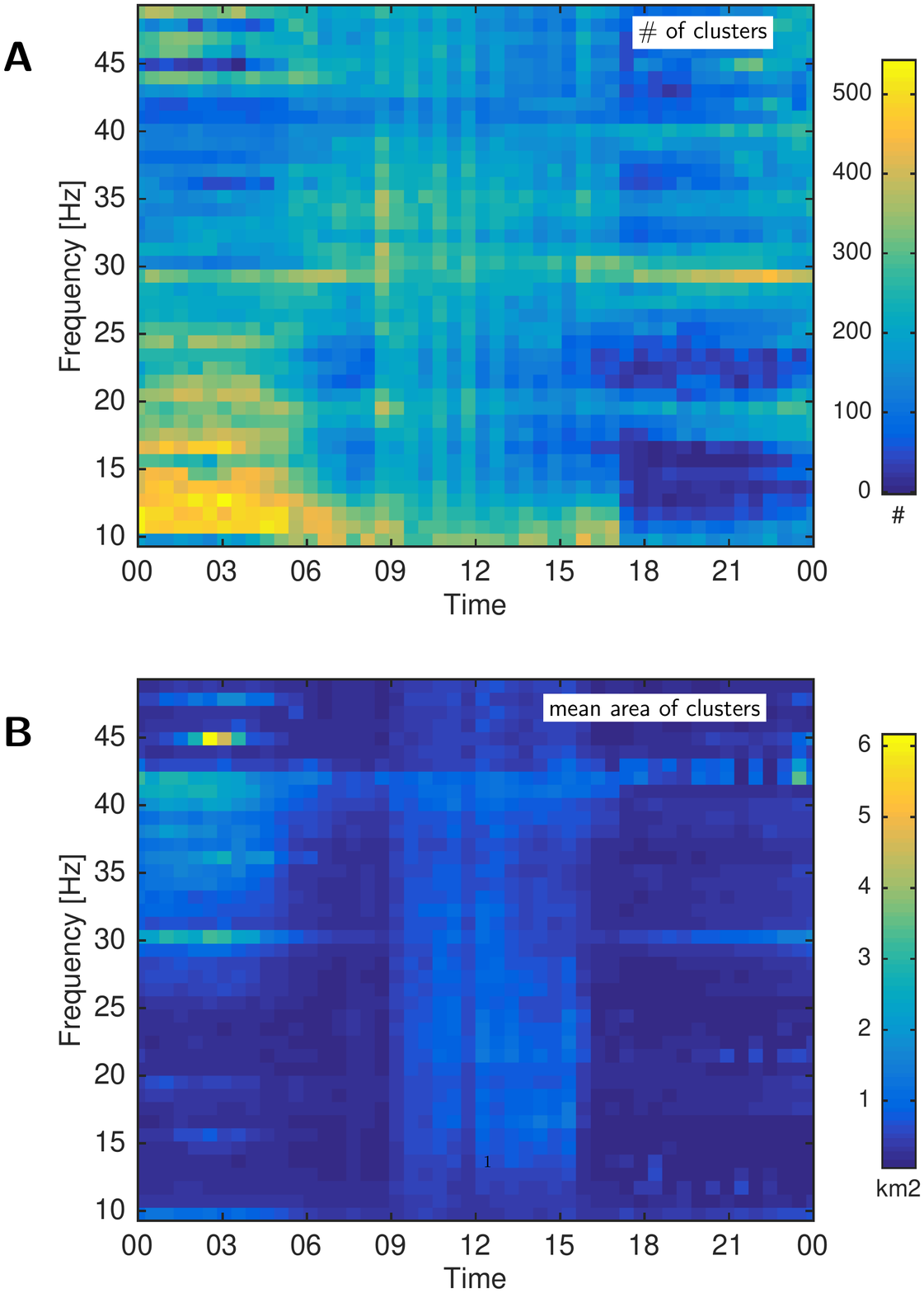}
\caption{
	Over 450,000 coherent sensor clusters with more than ten vertices were observed during the course of March 10th between 9.8\hbox{-}48.8~Hz. The number of clusters found in each hour and frequency bin is shown in {\em (A)}. The average spatial area covered by the convex hull of those clusters is shown in {\em (B)}. A wide-band increase in cluster area is seen from 9\hbox{-}15h and around 30--43~Hz in the early morning hours.
    }
\label{fig:area-stats}
\end{figure}

We perform the above analysis with the same parameterization for 41~frequency bins from 9.8\hbox{-}48.8~Hz for 24~hours starting on March~10th. For every detected coherent sensor cluster we store the mean of the coordinates (\ref{eq:spat-gauss}), the area of the convex hull around the cluster, and the frequency and time at which the cluster is observed. More than 450,000 clusters with more than ten vertices were observed during the course of March 10th in the given frequency bins. Figure~\ref{fig:area-stats}A shows the number of clusters found in each analysis hour and frequency bin and Figure~\ref{fig:area-stats}B gives the average area covered by those clusters. We find that in most frequency bands 90\% of clusters cover less than 1.5\% of the array aperture.

\begin{figure*}[t]
\center
\includegraphics[origin=c,width=1.6\columnwidth]{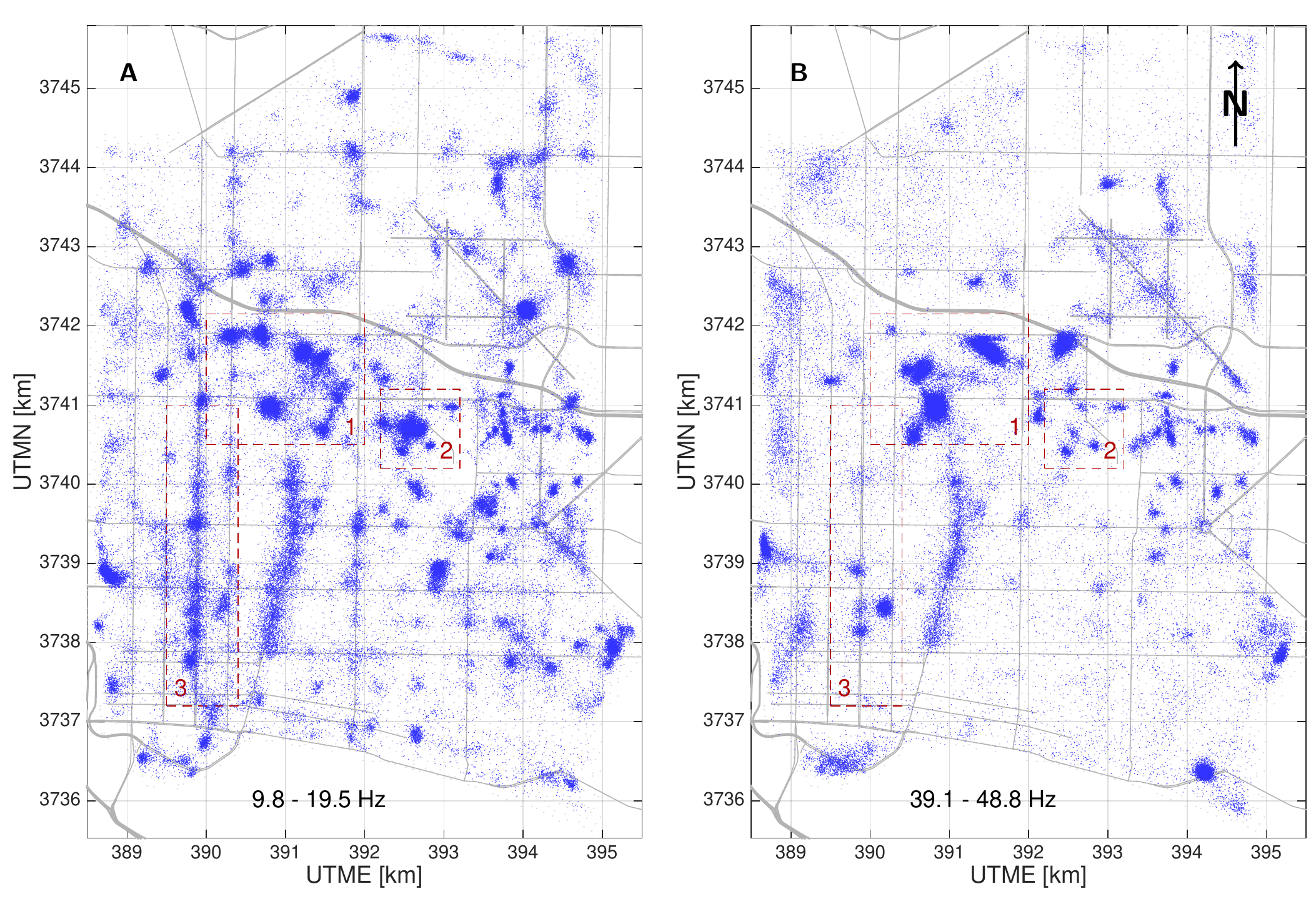}
\caption{
	Overview of all clusters identified during March 10th in the frequency range 9.8--19.5~Hz {\em (A)} and 39.1--48.8~Hz {\em (B)}. Each dot represents the center of a sensor cluster. Of those clusters 90\% cover less than 1.5\% of the array aperture area. Clusters in regions~1 and~2 coincide with oil production infrastructure. Region~1 contains several pump jacks and drill rigs while region~2 contains the central pump facility. Note how the spatial distribution of detections differs for the two bands. Region~3 highlights a strip of Long Beach Blvd that coincides with a section of the Long Beach light rail metro service. The strip is particularly active in low-frequencies. 
	    }
\label{fig:map-stats}
\end{figure*}

The maps in Figure~\ref{fig:map-stats} give a geographic overview of the identified cluster centers in the course of March 10th for  frequency bands 9.8--19.5~Hz and 39.1--48.8~Hz, each containing 10~frequency bins. A wealth of information could be collected in the results. Three particular regions with more sources are identified. Our analysis considers every frequency bin separately. This can be readily extended to aggregate clusters in different frequency bands that occur at a similar location and time.

\section{Conclusion}

We have introduced a source localization method for large aperture dense arrays that makes minimal assumptions about the propagation medium. The method requires that source signals exhibit significant strength only over a small distance within the array.
The spatial covariance matrix of a wave field has been reinterpreted as a matrix whose support is a connectivity matrix of a network of vertices (sensors) connected into communities.
These communities correspond to sensor clusters associated with individual sources.
The support of the covariance matrix has been estimated from limited-time data using a hypothesis test combined with a physical distance criterion. The latter was successful in preventing network communities from forming by chance

The method was tested on simulated data from a dense array containing three concurrent sources. From a total of 900~simulated sources only~3\% were not detected. A real data example from a 5200~element seismic array demonstrates the high degree of detail that the method can reveal about man-made noise sources within the array.


%

\appendices

%
%
%
%
%
%

\section{Some remarks about matrix and vector support}
\label{app:indicator-ops}

Define the support indicator function of a vector ${\bf v}\in \mathbb{C}^N$:
\begin{equation}
\hat{{\bf v}}=\mathcal{I}({\bf v}) \text{ where } \hat{{\bf v}}_i =
	\begin{cases}
               1               & \text{if } {\bf v}_i \ne 0  \\
               0 & \text{if } {\bf v}_i=0
           \end{cases}
\end{equation}
and analogously an indicator function $\mathcal{I}({\bf V})$ for a matrix ${\bf V}$. We then have:
%
\begin{align}
\mathcal{I}({\bf vw}^H) &= \hat{{\bf v}} \hat{{\bf w}}^T \\
\label{eq:indscale-ind}
\mathcal{I}(a{\bf v}) &= \hat{{\bf v}} \quad \quad \text{(for } a \ne 0 \text{)} \\
\label{eq:indicator-sum}
\mathcal{I}({\bf V}+{\bf W}) &= \hat{{\bf V}}+\hat{{\bf W}}
	- (\hat{{\bf V}} \circ \hat{{\bf W}}) \: ,
\end{align}
where $\circ$ is the Hadamard operator (element-wise multiplication). Assume that the supports of ${\bf v}$ and ${\bf w}$ do not overlap, i.e. $\hat{{\bf v}} \circ \hat{{\bf w}}={\bf 0}$, the zero-vector. Therefore
\begin{equation}
\label{eq:support-nooverlap}
(\hat{{\bf v}}\hat{{\bf v}}^T) \circ (\hat{{\bf w}}\hat{{\bf w}}^T )
	= \left( \hat{{\bf v}}\circ \hat{{\bf w}} \right) 
	\left( \hat{{\bf v}} \circ \hat{{\bf w}} \right)^T = {\bf 0}
\end{equation}
is the zero matrix and consequently, from (\ref{eq:indicator-sum}) and (\ref{eq:support-nooverlap})
\begin{align}
\mathcal{I}({\bf vv}^H + {\bf ww}^H) &= 
	 \hat{{\bf v}}\hat{{\bf v}}^T + \hat{{\bf w}}\hat{{\bf w}}^T 
		 - (\hat{{\bf v}}\hat{{\bf v}}^T \circ \hat{{\bf w}}\hat{{\bf w}}^T) \nonumber \\
	\label{eq:ind-sum-sum-ind}
	 &= \hat{{\bf v}}\hat{{\bf v}}^T + \hat{{\bf w}}\hat{{\bf w}}^T  \: .
\end{align}

In other words, if the support of vectors ${\bf v}$ and ${\bf w}$ do not overlap, then the support of the sum of their outer products is simply the sum of the outer products of their support-indicators.

\section{Connectivity Matrix operations}
\label{app:connectivity-supp}

Let $G$ be a network with $N$ vertices and let $A,B \subset G$ be two subsets of vertices that do not overlap, $A\cap B = \emptyset$. Define the indicator vector of a set of vertices $A$ as:
\begin{equation}
\label{eq:indicator-vec}
    {\bf a} = [a_1,\ldots,a_N]^T \text{ where } a_i =
    \begin{cases}
               1               & \text{if } i \in A  \\
               0 & \text{if } i \notin A
           \end{cases}
\end{equation}

Consider a network where $A$~is the only connected component. In this case
The connectivity matrix ${\bf E}$ for a network where $A$~is the only connected component has a support that is equal or less than ${\bf aa}^T$.

Let ${\bf E}$ be the connectivity matrix of a network~$G$. The following statements are equivalent:
\begin{enumerate}
	\item The network $G$ has two (and only two) connected components~$A$ and~$B$ that are not connected with each other, $A \cap B = \emptyset$.
	\item ${\bf E}$ has a support that is equal or smaller than ${\bf aa}^T + {\bf bb}^T$.
\end{enumerate}

{\em Proof:} Consider first the fully connected case, i.e. $a_i=a_j=1\: \forall i,j$ and likewise for ${\bf b}$. If the first item is true, then $a_ia_j=0$ if either $i \notin A$ or $j \notin A$. Also, $b_ib_j=0$ if either $i \notin B$ or $j \notin B$. Therefore we have $E_{ij}=a_ia_j+b_ib_j>0$ only if $i,j \in A$ or $i,j \in B$ and the second item is true.
The other direction follows a similar logic to argument just given.
Because the fully connected component contains all possible edges within a set of vertices any lesser connected components will have a subset of those edges, i.e. the connectivity matrix can only have fewer non-zero values. $\blacksquare$

Note that the union of the support of a connectivity matrix with the support of the unity matrix does not affect connected components that have a size larger than~one.

\section*{Acknowledgment}

We thank NodalSeismic LLC and Signal Hill Petroleum Inc. for sharing the data, in particular Dan Hollis and Robert Clayton for their efforts in making the data available. This work is supported by NSF grant no. EAR-0944109 and by the Swiss National Science Foundation.

\ifCLASSOPTIONcaptionsoff
  \newpage
\fi



\bibliographystyle{IEEEtran}
%
\bibliography{LITERATURENRI}

%
%

%


%




\vfill


\end{document}